\documentclass[mathleft
]{an}
\usepackage{graphicx}
\usepackage{rotating}
\usepackage{lscape}
\usepackage{longtable}
\usepackage{times}
\usepackage{fge}
\begin{document}

\Pagespan{789}{}
\Yearpublication{2011}%
\Yearsubmission{2010}%
\Month{11}%
\Volume{999}%
\Issue{88}%

\title{A transient event in AD 775 reported by al-\d{T}abar$\bar{\i}$: A bolide - not a nova, supernova, or kilonova}

\author{R. Neuh\"auser\inst{1} \thanks{Corresponding author: \email{rne@astro.uni-jena.de}}
\and P. Kunitzsch\inst{2}
}

\titlerunning{Transient celestial event in AD 775}
\authorrunning{Neuh\"auser \& Kunitzsch}

\institute{
Astrophysikalisches Institut und Universit\"ats-Sternwarte, FSU Jena,
Schillerg\"a\ss chen 2-3, D-07745 Jena, Germany
\and
LMU Munich, Germany (retired); home: Davidstrasse 17, 81927 M\"unchen, Germany
}

\received{12 June 2014}
\accepted{7 Aug 2014}
\publonline{ }


\abstract{Given that the cause for the strong increase in $^{14}$C in AD 774/5 in Japanese and German
trees is still a matter of debate (e.g. short Gamma-Ray Burst or solar super-flare), 
we have searched in Arabic chronicles for reports about unusual transient celestial events. 
In the {\em History of al-\d{T}abar}\textit{\={\i}} we found two (almost identical) reports about such an event. 
The group around caliph {\it al-Man\d{s}\={u}r} observed a transient event 
while on the way from Baghdad to Mecca on AD 775 Aug 29 - Sep 1 (Julian calendar).
A celestial object ({\em kawkab}) was seen to fall or set ({\em inqa$\d{d}\d{d}$a}), 
and its trace ({\em atharuhu}) was seen for at least tens of minutes (up to 70-90 min) during morning twilight.
The reports use the Arabic words {\em kawkab} and {\em athar(uhu)}, which were also used in the 
known Arabic reports about supernovae SN 1006 and 1054, so that one might consider an interpretation as a nova-like event.
The {\em kawkab} (celestial object) was observed only during the morning twilight at a brightness of
probably between about $-3$ and $0$ mag. Such a brightness and time-scale would be expected for
optical kilonovae (at $\sim 3$ to 9 kpc) in the context of short Gamma-Ray Bursts. There are no similar reports from eastern Asia
for this time. However, the short reports are fully consistent with a bolide: 
The word {\em kawkab} can be used for {\em meteor}, the verb {\em inqa$\d{d}\d{d}$a} normally means {\em falling down},  
the word {\em atharuhu} can mean {\em its trace}. We therefore prefer the interpretation as bolide.
We discuss in detail how to convert the Muslim calendar date to a date in the Julian calendar
using first the calculated Islamic calendar and then considering the time when the crescent new moon could
be visible at the given location.
}

\maketitle

\section{\it Introduction: $^{14}$C variation in AD 774/5}

Miyake et al. (2012) detected a significant variation in the isotope ratio 
of $^{14}$C to $^{12}$C in two Japanese trees around AD 774,
which was confirmed by Usoskin et al. (2013) with a German tree.
Depending on the carbon cycle model, the energy estimate would be either
$7 \times 10^{24}$ erg in $\gamma$- and/or cosmic rays at Earth within one year,
if the radioisotopes were formed due to incoming $\gamma$-photons above 10 MeV
with a supernova-like spectrum with a power-law index of $-2.5$ (Miyake et al. 2012),
or 4 to 6 times less using a different carbon cycle model (Usoskin et al. 2013).

Since there was significantly ($7.2~\sigma$) more $^{14}$C detected in the AD 775 trees
than in AD 774, the event should have taken place in AD 774.
However, there may be some uncertainties in the exact dating:
(i) Some trees may not form any rings at all in very cold years (e.g. Mann et al. 2012); 
(ii) some trees may be able to form two rings within one year, e.g. one in a warm early
spring and then, after a cold phase, one more in a warm fall 
(H. \"Ozbal and P. Kuniholm, priv. comm.); and
(iii) there are age offsets between the tree ages of different laboratories, 
e.g. $\pm$ a few yr by comparing Irish oak trees with Seattle measurements (e.g. Reimer et al. 2004).
Even though those effects can partly cancel each other, the exact year of the $^{14}$C increase 
and a possible event may be somewhat uncertain by up to a few years.
And even though the $^{14}$C variation in AD 774/5 was detected in several different trees
on different continents, always for the same year AD 774/5, that does not neccessarily
mean that the year is exactly correct (nor that, e.g., no tree rings are missing),
because it could be that in a cold year, no trees at all form a new ring,
e.g. due to a large vocano eruption with global effects, see Mann et al. 2012.

Given this timing uncertainty, it is neccessary to search for unusual highly energetic
events around the time of AD 774/5, and even if none were found, this would constrain
the cause of the $^{14}$C variation.

Miyake et al. (2012) argued that neither a supernova (SN) nor a solar or stellar flare
could be the cause for this event, because there is no Chinese observation
of a historic SN known around that time and because solar and stellar flares are neither strong
nor hard enough for the AD 774/5 variation in $^{14}$C and possibly also $^{10}$Be.

Allen (2012) suggested that a certain report in the Ang- lo-Saxon Chronicle
({\em This year also appeared in the heavens a red crucifix, after sunset}), 
presumably for AD 774, may be an absorbed SN; however, a SN 
cannot be observed as resolved or extended object ({\em cross}).
The sightings of a {\em red cross} or {\em crucifix} 
can easily be explained as halo display with horizontal arc and vertical pillar, possibly
also with sun-dogs (parhelion) or moon-dogs (paraselene), 
the frequent reports about them in medieval times can be explained by the fact that
the Christian monks (who reported them) thought at that time that the sighting of
a cross in the sky (or heaven) would indicate the presumable return of Jesus or their
Messiah, see Neuh\"auser \& Neuh\"auser (2014).
It was also shown that the above {\em red cross} was observed in AD 776, not AD 774/5 (e.g. Gibbons \& Werner 2012).

The nova or SN candidate listed as {\em Hye Sung} (an ancient Korean comet name) in Chu (1968) for
AD 776 Jun 1-30 in Tau-Aur (from the Korean Lee Dynasty chronology during the reign of He Gong Sinla)
is more likely a comet or nova (observed for only 30 days), possibly without a tail.

Hambaryan \& Neuh\"auser (2013, HN13) considered absorbed SNe quantitatively:
For the typical SN energy of $\sim 10^{51}$ erg, of which $\sim 1\%$ would be radiated in $\gamma$-rays, 
the hypothetical SN would have taken place at some $\sim 124$ pc only - for the Miyake et al. (2012)
energy estimate; for the lower energy estimate from Usoskin et al. (2013), the distance can be $\sqrt{4-6}$ times larger.
If the SN was either more energetic and/or more energy would have been radiated in $\gamma$-rays,
then the SN could have been further away. In the latter case, its SN remnant and possible 
compact remnant might have remained undetected so far.

Cliver et al. (2013) and Neuh\"auser \& Hambaryan (2014) both discussed the different
doubts that are connected with the solar super-flare hypothesis (Usoskin et al. 2013).
The suggestion by Liu et al. (2014), that a comet had impacted with Earth in AD 773 Jan bringing sufficient
$^{14}$C and $^{10}$Be, was falsified by Chapman et al. (2014), who could show that this comet was observed
as normal comet with a long tail in different nights from both China and Japan.

HN13 also showed that all observables of the AD 774/5 event are consistent with a short gamma-ray burst (GRB).
More recently, Pavlov et al. (2013 a,b) confirmed the approximate estimates by HN13 with more precise
calculations using GEANT and could show that a Galactic {\em long} GRB is also not inconsistent with
the AD 774/5 observables. Given the energy estimate by Miyake et al. (2012),
HN13 estimated the distance towards the short GRB to be $\sim 0.1$ to $\sim 4$ kpc
(again, $\sqrt{4-6}$ times more distant for the Usoskin et al. (2013) energy estimate,
i.e. up to $\sim 9$ kpc including the Galactic Center) -
but probably outside of 1 kpc, because there was no mass extinction event on Earth around AD 774. 
A short GRB is expected to be due to the merger of two compact objects
like black holes or neutron stars, possibly also white dwarfs.
The occurence rates of short GRBs and compact mergers are highly uncertain, possibly very low (see HN13).

A radioactively powered optical (or near infrared) transient is expected (so-called kilo-nova) from compact mergers 
$M_{\rm V} = -15$ mag at peak (Metzger \& Berger 2012; Piran, Nakar, Rosswog 2013; Grossman et al. 2014).
This corresponds to $m_{\rm V} = -10$ mag at 0.1 kpc, $-5$ mag at 1 kpc, $-2$ mag at 4 kpc 
(or $m_{\rm V} =-0.2$ mag for 9 kpc), all values for negligible absorption (otherwise fainter).
Hence, it may have been observable by naked eye, but only for a short duration between a few hours
and a few days (Metzger \& Berger 2012; Piran, Nakar, Rosswog 2013; Grossman et al. 2014), 
i.e. much shorter than a typical SN or nova. An extra-galactic kilo-nova may once have been observed
after the short Gamma-ray burst GRB~130603~B (Tanvir et al. 2013).

This motivated us to search for reports in old chronicles about an unusual transient celestial event.
Given that European and eastern Asian (Chinese, Korean, and Japanese) chronicles on astronomical events,
even for the 8th century, were fully studied before (see e.g. Stephenson \& Green 2002 and
references therein), we concentrated on Arabic chronicles.

Here, we present a transient celestial event reported by the Persian historian {\it al-\d{T}abar}\textit{\={\i}};
it could have been either a bolide or a nova-like event.\footnote{While we were writing this manuscript,
the Arabic text given below from {\it al-\d{T}abar}\textit{\={\i}} was independently also found 
and considered to be relevant by Jalal Hojati Fahim from Iran (priv. comm.).}
In Sect. 2, we present the Arabic text and its translation to English.
In Sect. 3, we discuss the circumstances of the sighting: Location, observers, sources, and dating.
The interpretation by the observers is than given in Sect. 4.
Then, in Sect. 5, we present an interpretation as bolide.
In Sect. 6, we consider a speculative interpretation as nova-like event.
Finally, we summarize the results in Sect. 7.

\section{{\it The Arabic text by} {\it al-\d{T}abar}\textit{\={\i}}}

The well-known Persian historian named 
{\it Ab\={u} Ja$^{\rm c}$far Mu\d{h}a- mmad ibn Jar}\textit{\={\i}}{\it r al-\d{T}abar}\textit{\={\i}} (short: {\it al-\d{T}abar}\textit{\={\i}})
was born AD 839/840 in Amul (today in Iran) and died AD 923 in Baghdad (Iraq). Among other subjects,
he studied Islamic religion, mathematics, and medicine. One of his most influential and best known works
is the historical chronicle {\em T\={a}r}\textit{\={\i}}{\em kh al-Rusul wa-al-Mul\={u}k} (History of the Prophets and Kings), 
often referred to as {\em History of al-\d{T}abar}\textit{\={\i}}. This chronicle giv- es short reports for all years 
for the 8th and 9th centuries, with more details in some years,
in particular for years, when a caliph died, e.g. the Islamic year 158 hijra (158h)\footnote{The year 158 hijra (158h) is 158 lunar years after the start of the lunar year 
in which the Hijra took place, 
i.e. the emigration of the Islamic Prophet Mu\d{h}ammad from Mecca to Medina, known as Hijra;
this era, i.e. the year 1h started on AD 622 Jul 16 according to most scholars - but it may have been on AD 622 Jul 15 according
to, e.g., de Blois (2000); according to NASA GSFC (eclipse.gsfc.nasa.gov/phases/), new moon was on AD 622 Jul 14 (Julian calendar) 
at 5:26h UT ($\pm 2$ min according to Morrison \& Stephenson 2004), so that the crescent new moon was hardly visible 
in the evening of AD 622 Jul 15, but it was well visible in Mecca, Saudi Arabia, in the evening of AD 622 Jul 16 (Neugebauer 1929).}, 
which runs from AD 774 Nov to AD 775 Oct.
In most cases, {\it al-\d{T}abar}\textit{\={\i}} quotes and mentions the source(s) of his reports.
See, e.g., Kennedy (1990) and Sezgin (1967) for more details about {\it al-\d{T}abar}\textit{\={\i}} and his chronicle.

{\it Al-\d{T}abar}\textit{\={\i}} reports briefly about the year 157h (running from AD 773 Nov to AD 774 Nov),
for which he does not report any celestial or otherwise peculiar event.
The report about the next year (158h) is much more detailed,
because the caliph {\it Ab\={u} Ja$^{\rm c}$far $^{\rm c}$Abdall\={a}h ibn Mu\d{h}ammad al-Man\d{s}\={u}r} 
(short: {\it Ab\={u} Ja$^{\rm c}$far} or {\it al-Man\d{s}\={u}r}) died in that year.
{\it Al-\d{T}abar}\textit{\={\i}} reports in detail the events in the weeks before and after his death.

As part of the events leading to the death of caliph {\it al-Man\d{s}\={u}r}, {\it al-\d{T}abar}\textit{\={\i}} reports at two
occasions of his chronicle about a peculiar transient celestial event. The two Arabic reports in {\it al-\d{T}abar}\textit{\={\i}} are almost identical.

The Arabic texts are shown in Figs. 1 and 2 from the Leiden edition of the {\em History of al-\d{T}abar}\textit{\={\i}} (de Goeje 1879).

\begin{figure*}
{\includegraphics[angle=0,width=17cm]{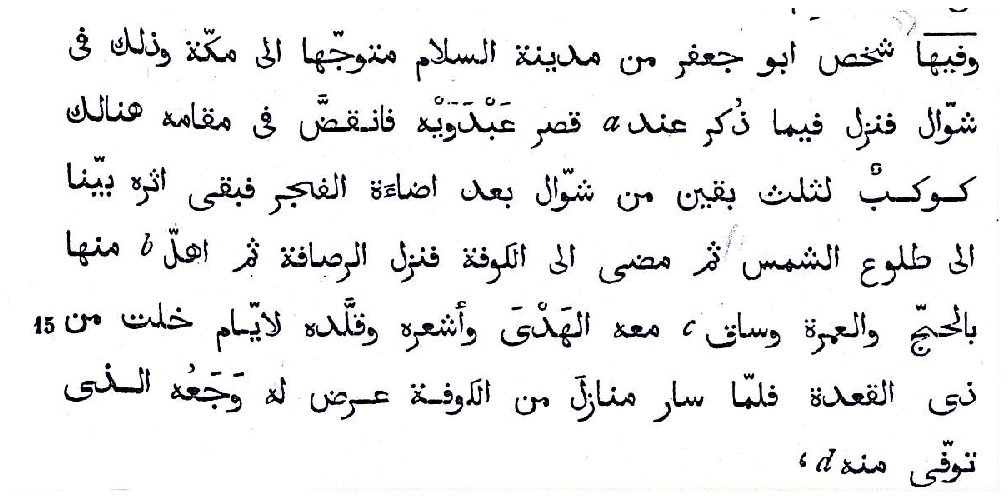}}
\caption{A copy of a small part of the {\em History of al-\d{T}abar}\textit{\={\i}} in Arabic, here one paragraph from the 
reports about year 158h (AD 774/5) with the text about the transient celestial event.
This is what we call the 1st report, located within a report about the illness and death of
Caliph {\it Ab\={u} Ja$^{\rm c}$far al-Man\d{s}\={u}r}. The word {\em kawkab} is the first word (to the right) in the 3rd line
(from top), and the word {\em atharuhu} is seen as 2nd-to-last word in the 3rd line (2nd from left).
Comment (a) after the word $^{\rm c}${\em inda} says that, in one of the manuscripts (ms B), this word is omitted,
which does not change the meaning; comments b, c, and d are after the relevant part about the celestial event.
See Sect. 2 for the Arabic text in English transcription and its English translation.}
\end{figure*}

\begin{figure*}
{\includegraphics[angle=0,width=17cm]{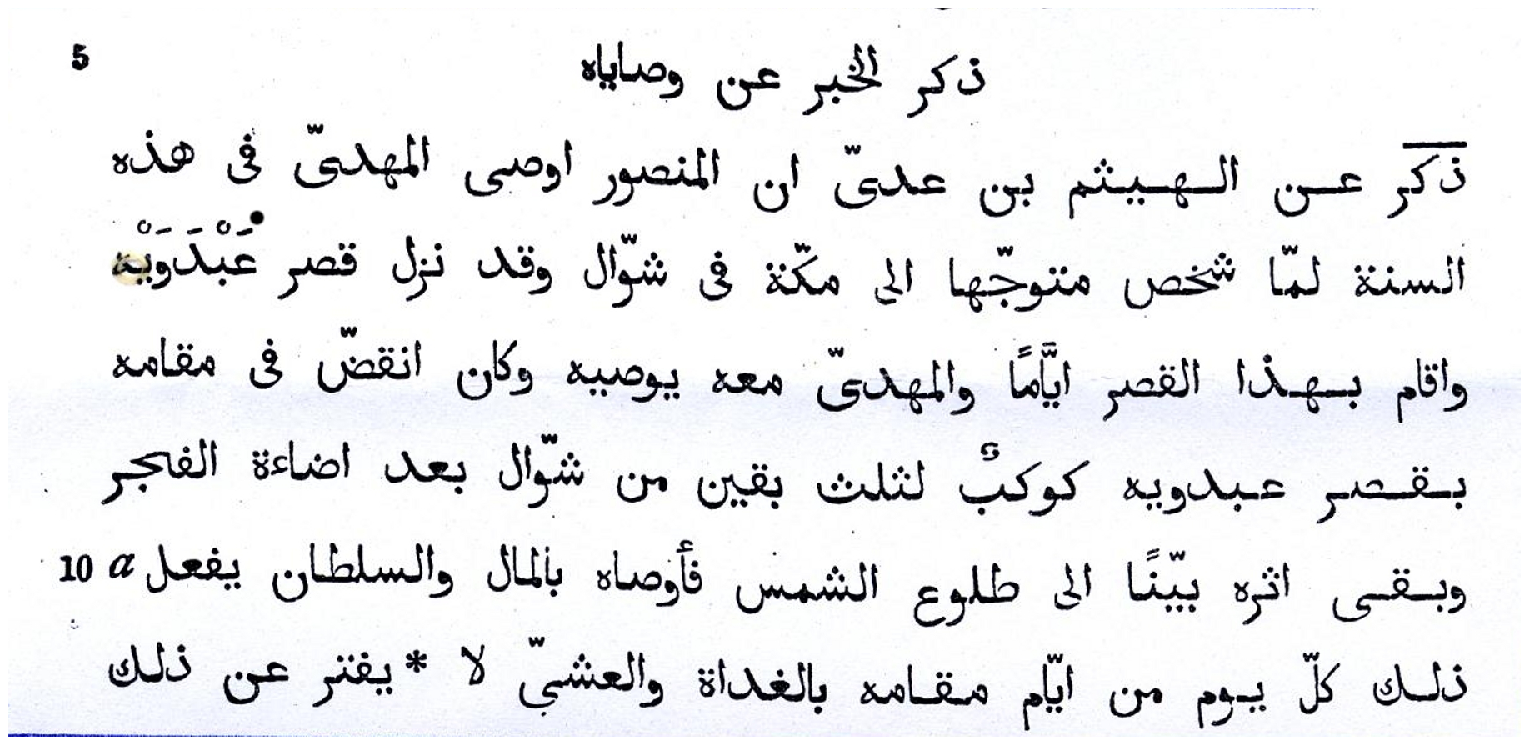}}
\caption{A copy of a small part of the {\em History of al-\d{T}abar}\textit{\={\i}} in Arabic, again one paragraph from the 
reports about year 158h (AD 774/5) with the text about the transient celestial event.
This is what we call the 2nd report, located at the beginning of a report about the last will
of caliph {\it Ab\={u} Ja$^{\rm c}$far al-Man\d{s}\={u}r}. 
This part starts with the heading line (centered uppermost line), 
then the start of the report about the caliph's last will. 
The text about the celestial event starts in the 4th line
from the bottom almost at the end of the line (at the left) with {\em wa k\={a}na ...}. The word
{\em kawkab} is the 3rd word (from the right) in the 3rd-to-last line and the word
{\em atharuhu} is seen as 2nd word (from the right) in the 2nd-to-last line.
(Comment (a) comes after the report about the celestial event.)
See Sect. 2 for the Arabic text in English transcription and its English translation.}
\end{figure*}

The first appearance of the celestial event is mentioned within a report about {\it al-Man\d{s}\={u}r}`s illness, 
which was probably the cause for his death. The Arabic text reads as follows (Fig. 1, lines 2-4): \\
{\it fa-nazala f}\textit{\={\i}}{\it m\={a} dhukira $^{\rm c}$inda Qa$\d{s}$r $^{\rm c}$Abdawayh
fa-nqa$\d{d}\d{d}$a f}\textit{\={\i}} {\it muq\={a}mihi hun\={a}lika kawkab
li-thal\={a}th baq}\textit{\={\i}}{\it na min Shaww\={a}l
ba$^{\rm c}$da i$\d{d}$\={a}$\,\spirituslenis{^{ }}$at al-fajr fa-baqiya atharuhu bayyinan il\={a} $\d{t}$ul\={u}$^{\rm c}$ \\
al-shams.}

Our own English translation runs as follows: \\
{\em He [caliph al-Man\d{s}\={u}r] rested - as has been reported - near Qa\d{s}r~~$^{\rm c}$Abdawayh.
During his stay there, there fell down a star [kawkab] in the third last night of Shaww\={a}l
after the beginning of dawn; its trace [atharuhu] remained visible until sunrise.}

A previous translation of this text by Kennedy (1990) is as follows (together with context, his pages 87-89,
dates as given by him): \\
{\em According to $^{\rm c}$Umar b. Shabbah - Mu\d{h}ammad b. $^{\rm c}$Imr\={a}n, freedman of Mu\d{h}ammad b.
Ibr\={a}h}\textit{\={\i}}{\it m b. Mu\d{h}ammad b. $^{\rm c}$Al}\textit{\={\i}} {\it b.
$^{\rm c}$Abdall\={a}h b. $^{\rm c}$Abb\={a}s - his father: 
Al-Man\d{s}\={u}r wrote to Mu\d{h}ammad b. Ibr\={a}h}\textit{\={\i}}{\it m, governor of Mecca ... 
When Ab\={u} Ja$^{\rm c}$far mounted [his camel] and he and his companion al-Rab}\textit{\={\i}}{\it $^{\rm c}$ 
went on, Mu\d{h}ammad [b. Ibr\={a}h}\textit{\={\i}}{\it m] ordered his doctor to go over
to Ab\={u} Ja$^{\rm c}$far's camel-kneeling place, and he saw his 
[Ab\={u} Ja$^{\rm c}$far's] excrement and he said to Mu\d{h}ammad, $^{\prime \prime}$I saw the excrement
of a man who does not have long to live$^{\prime \prime}$. When he [Ab\={u} Ja$^{\rm c}$far] entered Mecca 
he remained there only a short time before he died and Mu\d{h}ammad was saved.
In Shaww\={a}l of this year (AD 775 Aug 4 - Sep 1) Ab\={u} Ja$^{\rm c}$far set out 
from the City of Peace [Baghdad] heading for Mecca.
Is is said that he stayed at Qa\d{s}r~~$^{\rm c}$Abdawayh and 
while he was there a meteorite [kawkab] fell [inqa\d{d}\d{d}a] on 26 Shaww\={a}l (Aug 29) after 
the beginning of dawn and remained visible until sunrise; then he went on to al-K\={u}fah and stayed at 
al-Ru\d{s}\={a}fah and then set
out from there on the pilgrimage and the lesser pilgrimage ... When he had gone some stages from al-K\={u}fah the
disease of which he died became apparent ... and then he died at dawn or sunrise of 
Saturday 6 Dh\={u} al-\d{H}ijjah (AD 775 Oct 7).}

This text tells us first of all the source: {\it Al-\d{T}abar}\textit{\={\i}} obtained the story 
from {\it $^{\rm c}$Umar b. Shabbah}, 
who in turn knew it from {\it Mu\d{h}ammad b. $^{\rm c}$Imr\={a}n}, 
a freedman of {\it Mu\d{h}ammad b. Ibr\={a}h}\textit{\={\i}}{\it m}, governor of Mecca, who knew the story from his father.
The text also tells us that this {\it Mu\d{h}ammad b. Ibr\={a}h}\textit{\={\i}}{\it m} as well as a doctor were members of the group around the caliph 
who saw the celestial event. According to the line of sources given by {\it al-\d{T}abar}\textit{\={\i}} (from {\it $^{\rm c}$Umar b. Shabbah}), 
it seems that the father of governor {\it Mu\d{h}ammad b. Ibr\={a}h}\textit{\={\i}}{\it m} 
was also a member of the group, and he saw and reported about the celestial event. \\
The quotation from the doctor may indicate blood in the excrement of the
caliph, possibly an indication of fatal cancer. Then, the sighting of the celestial event is reported, afterwards further details
about his illness and death. It is not clear nor explicitly mentioned, whether the celestial event is brought into (astrological) context
with the illness and death of the caliph. We will discuss the celestial event, its circumstances, and its interpretation
by the caliph in more detail below.

This particular observation was given in Cook (1999) as meteor using the translation by Kennedy (1990) as {\em falling star};
Cook (1999) gave AD 774 Aug 29 as date, i.e. wrong by about one year.
Cook (1999) also did not mention the second report in {\it al-\d{T}abar}\textit{\={\i}}, see below.
Furthermore, since Cook (1999) did not discuss the possible meanings of {\it kaw- kab}, {\it atharuhu}, and {\it inqa$\d{d}\d{d}$a},
which appear in both the first and second report (the latter was translated 
as {\em setting} by Dietrich 1952), it is important to discuss this event in detail.

The second report about the celestial event comes several pages later, when {\it al-\d{T}abar}\textit{\={\i}} reports 
how {\it al-Man\d{s}\={u}r} ga- ve his political testament (in oral form) to his son {\it al-Mahd}\textit{\={\i}}: 
As mentioned before, in the last weeks of the life of {\it al-Man\d{s}\={u}r}, the caliph travelled to Mecca;
on the first part of this journey, he was accompanied by his son and successor {\it al-Mahd}\textit{\={\i}}.
The chapter heading is {\em Information about his Wills}. At the beginning of this section, 
there is the second report about the transient celestial event, the text reads as follows (Fig. 2, lines 4-6): \\
{\it wa-k\={a}na inqa$\d{d}\d{d}$a f}\textit{\={\i}} {\it muq\={a}mihi bi-Qa$\d{s}$r $^{\rm c}$Abdawayh kawkab 
li-thal\={a}th baq}\textit{\={\i}}{\it na min Shaww\={a}l ba$^{\rm c}$da i\d{d}$\bar{a}\,\spirituslenis{^{ }}$at al-fajr 
wa-baq- iya atharuhu bayyinan il\={a} $\d{t}$ul\={u}$^{\rm c}$ al-shams.}

Our own English translation runs as follows: \\
{\em During his stay in Qa$\d{s}$r $^{\rm c}$Abdawayh there had fallen down a star [kawkab] in the
third last night of Shaww\={a}l after the beginning of dawn; its trace [atharuhu]
remained visible until sunrise.}

We would like to note that two other translations had been published before. 
Kennedy (1990) wrote (again together with its context, his pages 149/150, dates as given by him): \\
{\em Information about his Wills. According to al-Haytham ibn $^{\rm c}$Ad}\textit{\={\i}}{\it : Al-Man\d{s}\={u}r made a 
will for al-Mahd}\textit{\={\i}} {\it when he set out for Mecca in Shaww\={a}l of this year (AD 775 Aug 4 - Sep 1). He had stopped at Qa\d{s}r~~$^{\rm c}$Abdawayh 
and stayed some days in the castle there. 
Al-Mahd}\textit{\={\i}} {\it was with him, and he gave him his testament. While he was staying in Qa\d{s}r~~$^{\rm c}$Abdawayh, 
a shooting star [kawkab] fell [inqa\d{d}\d{d}a] on 28 Shaww\={a}l (AD 775 Aug 31) after it began to get light,
and its track [atharuhu] remained clear until sunrise. 
He [al-Man\d{s}\={u}r] bequeathed him [al-Mahd}\textit{\={\i}}{\it ] his wealth and authority (sul- \d{t}\={a}n) ...} \\
The source of the report is given here as {\em al-Haytham ibn $^{\rm c}$Ad}\textit{\={\i}}, a well-known historian
(birth AD 747, died AD 821 or 822), but no books written by him were found yet (Sezgin 1967, Leder 1991);
the source of {\it al-\d{T}abar}\textit{\={\i}} and others, who quote or depend on him, are lost books
and/or oral reports and teachings (Leder 1991, Seidensticker 1994).

Dietrich (1952) translated the relevant part about the celestial event as follows to German
(date given by him): \\
{\em W\"ahrend seines Aufenthaltes in Qa\d{s}r~~$^{\rm c}$Abdawayh war am 28. \v{S}auw\={a}l 
nach dem Aufleuchten der Morgenr\"ote ein
Kom- et niedergegangen, dessen Spur bis Sonnenaufgang sichtbar blieb.} \\
This German text is to be translated to English as follows: \\
{\em During his stay in Qa\d{s}r~~$^{\rm c}$Abdawayh, on 28th Shaww\={a}l, after the start of the morning twilight, 
a comet [kawkab] had set, whose trace [atharuhu] remained visible until sunrise.}

As mentioned in Cook (1999) for the first report by {\it al-\d{T}abar}\textit{\={\i}}, a report about this event can also be found
in the work by {\it Ibn al-Djawz}\textit{\={\i}} called {\em al-Munta\d{z}am f\textit{\={\i}} al-t\={a}r}\textit{\={\i}}{\em kh},
quo- ting page 203 in Vol. 8 in a 1992 edition published in Beirut by Alam Al-Kutub (Cook 1999). 
We could not find this text from {\it Ibn al-Djawz}\textit{\={\i}};
according to Laoust (1979), only the last six volumes from {\it Ibn al-Djawz}\textit{\={\i}}'s book 
{\em al-Munta\d{z}am f\textit{\={\i}} al-t\={a}r}\textit{\={\i}}{\em kh}
were found so far, and they start with the year 257h = AD 871 (and end in 574h = AD 1179).
Since {\it Ibn al-Djawz}\textit{\={\i}} lived
later (12th/13th century AD) than {\it al-\d{T}abar}\textit{\={\i}} (died AD 923), we cannot expect to find new information
in the report by {\it Ibn al-Djawz}\textit{\={\i}}.
Furthermore, Cook (1999) remarks that this event or a different report about it might have been listed also in Dall'Olmo (1978) for AD 776,
but the event given in Dall'Olmo (1978) on {\em two inflamed shields} (seen above a church in Germany during the day) 
presumably from Annales Bertiniani\footnote{Since the Annales Bertiniani, considered to be a continuation of the Royal Frankish Annales, 
run from AD 830 to 882, this report cannot be found in Bertiniani, but in the Royal Frankish Annales themselves; the Annales Bertiniani
were written by the Spanish bishop Prudentius of Troyes in France, who reports as eyewitness from AD 835 until his death in AD 861;
the author of the first few years since AD 830 is unknown; the 3rd part from AD 861-882 was written by Hinkmar according to a source 
from the next century; a copy of the text was first found in the monastery of St. Bertin, hence its name (Rau 1958).} were
two sun-dogs (Neuh\"au- ser \& Neuh\"auser 2014) mis-interpreted already by Dall'Ol- mo (1978) as {\em two bright meteors or an aurora}.

The two reports by {\it al-\d{T}abar}\textit{\={\i}} as cited above are almost identical 
except that the 1st report says {\em there fell down a star},
while the 2nd report has {\em there had fallen down a star}. 
Both reports give the same (Muslim) date (3rd-to-last night of {\em Shaww\={a}l}),
even though Kennedy (1990) gives two different (western) dates - clearly in AD 775,
which is well-established due to the death of the caliph in the same year slightly later.

\section{\it Location and Date}

The place {\em Qa\d{s}r~~$^{\rm c}$Abdawayh} 
is a castle or camp in the Baghdad area (Lassner 1970, Kennedy 1990).
The word is of Persian origin.
In the first report, it was said that, after leaving {\em Qa\d{s}r~~$^{\rm c}$Abdawayh}, 
the group then went to {\it al-K\={u}fah} and
stayed at {\it al-Ru\d{s}\={a}fah} (near {\it al-K\={u}fah}). 
{\it Al-K\={u}fah} is known to be located 170 km south of Baghdad on the
Euphrates river founded in AD 637/8, and {\it al-Ru\d{s}\={a}fah} is said to be close to {\it al-K\={u}fah}
(hence, probably not the part of Baghdad called {\it al-Ru\d{s}\={a}fah}).

The event was observed by {\it al-Man\d{s}\={u}r} and/or one or more members of the group
travelling with him to Mecca. In the first report, it is mentioned that the source of {\it al-\d{T}abar}\textit{\={\i}} was
connected to {\it Mu\d{h}ammad b. Ibr\={a}h}\textit{\={\i}}{\it m}, the govenour of Mecca, who was a member of the group. It is therefore
well possible that the report goes back to him. It is possible that also the historian 
{\it al-Haytham ibn $^{\rm c}$Ad}\textit{\={\i}},
mentioned as source of the second report, knew the story from the very same source.
If {\it al-\d{T}abar}\textit{\={\i}} used two apparently different sources, they may not have been independent of each other,
which could explain the almost identical wording in both reports.
At the beginning of the 2nd report, {\it al-\d{T}abar}\textit{\={\i}} starts the report with 
{\em Dhukira $^{\rm c}$an al-Haytham ibn $^{\rm c}$Ad}\textit{\={\i}} ...,
which means {\em as has been reported from al-Haytham ibn $^{\rm c}$Ad}\textit{\={\i}} ..., 
possibly indicating that the source was not
available to {\it al-\d{T}abar}\textit{\={\i}} in written form.
According to Leder (1991), who has studied all known reports citing {\it al-Haytham ibn $^{\rm c}$Ad}\textit{\={\i}} as source,
there is no further report (based on {\it al-Haytham ibn $^{\rm c}$Ad}\textit{\={\i}}) known about this particular celestial event.

Let us now discuss in detail the dating of the event. \\
In both reports, the date of the event is given as {\em in the third last night of Shaww\={a}l}. \\
The Muslim calendar is a lunar calendar.\footnote{In the Quran, Sura 2, 189 it is said
(as translated by Asad 1980, our additions in round brackets (), his comments in square brackets []):
{\em They will ask thee (Prophet Mu\d{h}ammad) about the new moons (Arabic: {\it hil\={a}l}, for English crescent of the new moon).
Say: They indicate the periods for [various doings of] mankind, including the pilgrimage.}
[The reference, at this stage, to lunar months arises from the fact that the observance of
several of the religious obligations instituted by Islam - like the fasting in {\em Rama\d{d}\={a}n}, or the
pilgrimage to Mecca ... - is based on the lunar calendar ...]}
A Muslim month runs from the sunset at the evening with the first sighting of the
crescent (new) moon (as confirmed by two independent observers) until the sunset with the next such sighting (one month later), 
the Muslim day runs from sunset to sunset.
See Ilyas (1994) for details about the Lunar Date line.
According to the Muslim Hadith tradition, a month in the Muslim calendar (or at least the month of {\em Rama\d{d}\={a}n}) should last 
either 29 or 30 days.\footnote{According
to the collection of Hadith by Bukh\={a}r\={\i} translated by M. Muhsin Khan (www.searchtruth.com/hadith$\_$books.php),
Prophet Mu\d{h}ammad once said: {\em When you see the crescent (of the month of Rama\d{d}\={a}n),
start fasting, and when you see the crescent (of the month of {\it Shaww\={a}l}), stop fasting; and if the sky is overcast (and you can't see it)
then regard the crescent (month) of Rama\d{d}\={a}n (as of 30 days).} }
See, e.g., Spuler \& Mayr (1961), 
Ilyas (1994),
de Blois (2000), and Said et al. (1989) for more details about Muslim calendar rules. \\
In the theoretical {\em calculated} Islamic calendar (not used in practice, at least not in former times), 
the months have an alternating (but otherwise fixed, except the leap month) length of 29 and 30 days,
e.g. {\it Shaww\={a}l} would then always have 29 days; then, 11 years with one leap day each were needed in 30 years;
starting with the day of the Hijra, one can then calculate the conversion between the Islamic calendar and each other calendar.
There are two basic uncertainties in the calculated Islamic calendar (Ginzel 1906, Spuler \& Mayr 1961):
The date of the start of the era, the Hijra, is not certain, probably on AD 622 Jul 15, but possibly on AD 622 Jul 14;
it is not clear whether the 15th or 16th year within an intervall of 30 years with 11 leap years would have been picked
for inserting one leap day at the end of the last month.
Hence, this calendar can deviate by up to two days (Ginzel 1906, Spuler \& Mayr 1961, de Blois 2000).  \\
In addition, it is quite clear that such a calculated Islamic calendar was not in use in historic times,
as the famous Arab astronomer {\it al-B\={\i}r\={u}n\={\i}} (AD 973-1048) wrote: {\em the leap years of the Arabs, 
not that the Arabs ever actually used or use them, but the authors of astronomical tables need them when they 
construct tables on the basis of the years of the Arabs} (quoted after de Blois 2000).
Hence, in reality, until recently,
the start (and end) of a month was set by {\em observing} the crescent new moon. \\
However, the sighting of a crescent new moon could have been delayed by one or more days due to, 
e.g., bad weather, high extinction at low latitude, or difficult landscape.
For that particular month ({\it Shaww\={a}l} 158h), no calibration is kno- wn, 
e.g. no mentioning of a week day together with a date or an astronomical event (such as, e.g., an eclipse)
with well-known precise date;
a calibration point outside of that month, e.g. the day of the death of the caliph,
may not be relevant, because such shifts (due to a delayed crescent sighting) were not neccessarily transferred
to the next month nor cumulative, but could be corrected with the next new crescent.
(If the length of a month would be fixed to, e.g., at least 29 days and/or at most 30 days,
then a calibration in a neighbouring month and/or the knowledge of the start and/or end of a 
neighbouring month would also help, see at the end of this section.)
However, there were some possibilities to find the start of the month even in case
of bad weather: One can count the number of days since the last start of a month
(knowing that the length of the lunar month is 29 or 30 days);
given that the length of the synodic lunar month is not a multiple of days and can even vary from 29.26 to 29.80 days,
this can introduce an error source of $\pm 1$ day.
Alternatively, one can observe the moon later and determine its age or phase.
With the latter method, one can correct the start of the month and the current date (inside a lunar month)
any time during the month. However, the precision of this method is also not perfect,
but has an error of at least $\pm 1$ day, if not $\pm 2$ days.
In addition, some $15~\%$ of claimed lunar crescent sightings even by experienced observers can be wrong,
e.g. being reported several hours too early (Doggett \& Schaefer 1994).
Hence, without a calibration point in the given month (e.g. an independent dating by a different source in a different calendar
or a weekday mentioned together with a date), any conversion between a Muslim date and a Julian/Gregorian date has
an error of $\pm 1$ day ($\sim 1 \sigma$ error bar), i.e. it is sometimes even 2 days off. \\
According to the calculated Islamic calendar, though, the month of {\it Shaww\={a}l} in the year 158h
would be running from AD 775 Aug 4 to Sep 1 (according to, e.g., Spuler \& Mayr 1961).
To convert a Muslim date to a Julian or Gregorian date, the calculated Islamic calendar can
be used as a first approximation, but if an accuracy of better than one or two days is needed
and/or if the precision or error bar is needed, one should consider the exact time of crescent 
new moon sighting at the relevant location, which we will do next. \\
According to NASA GSFC\footnote{eclipse.gsfc.nasa.gov}, new moon was on AD 775 Aug 1 (Julian calendar) at 6:57h UT ($\pm 53$ min),
according to Gautschy (2011)\footnote{www.gautschy.ch/~rita/archast/mond/Babylonerste.txt}, it was at 6:59h UT.
According to Gautschy (2014), the crescent new moon can be seen if the lag-time between sunset and moonset and the difference in
their azimut are large enough, see Fig. 3, weather permitting.
The earliest reported sighting of the crescent (new) moon was 15.4 hours (by naked-eye viewing by Julius Sch- midt, 
a modern 19th century observer in Greece)
after the exact new moon time (Fotheringham 1910, Schaefer 1993), while most sightings are even later.
Hence, the month of {\it Shaww\={a}l} 158h should have started with the first sighting of the crescent in the evening on Aug 2 or 3, see Fig. 3;
the crescent new moon was theoretically visible on Aug 2, but if the weather was bad, the month of {\em Rama\d{d}\={a}n} ended one day later on Aug 3. 
According to Gautschy (2011) (footnote 7), the crescent new moon of this particular month was not visible before AD 775 Aug 2 at 16:23h UT at Babylon
(not far from Baghdad, Iraq), actually visible at the first dawn after the time given. 
Hence, this month probably started on Aug 2 (in case of good weather), but maybe one day later on Aug 3 -
or possibly on Aug 1 (in case of a false early sighting). \\
The next new moon was 29.5 days later on AD 775 Aug 30 at 22:44h UT ($\pm 2$ min 
according to Morrison \& Stephenson 2004), so that the end of the month of {\it Shaww\={a}l}
(and the start of the next month, called {\it Dh\={u} al-Qa$^{\rm c}$dah}) should have been on AD 775 Sep 1 or 2, while the crescent new moon
was still too low and too close to the Sun on Aug 31, see Fig. 3. According to Gautschy (2011) (footnote 7), the crescent new moon of this 
particular month was not visible before AD 775 Sep 1 at 15:49h UT at Babylon, actually visible at the first dawn after the time given.
These dates are Julian calendar dates (the difference between the old Julian calendar and the new Gregorian calendar was 10 days in AD 1582,
so that, by the time of AD 775, the shift was three days). Hence, the month of {\it Shaww\={a}l} ended probably on Sep 1, but maybe one
day later (in case of bad weather on Sep 1) or possibly on Aug 31 (in case of a false early sighting). \\
We can therefore conclude that the month of {\it Shaww\={a}l} 158h started
with the sighting of the crescent (new) moon on AD 775 Aug $2 \pm 1$ (Julian calendar). 
The end of this month was then shortly after the next new moon
(AD 775 Aug 30, 22:44h UT, $\pm 2$ min according to Morrison \& Stephenson 2004), 
i.e. probably at sunset on AD 775 Sep 1 ($\pm 1$ day) in the Julian calendar. \\
According to de Blois (2000), the wording {\it li-thal\={a}th baq}\textit{\={\i}}{\it na min Shaww\={a}l}
({\it third-to-last-night of Shaww\={a}l}) means the 27th of {\it Shaww\={a}l}, as all such dates
were given and memorized under the assumption that the month would have a total of 30 days.
One could question this by arguing that, after it was noticed that a particular month had only
29 days, one could correct the dating; in such a case, the third-to-last-night of {\it Shaww\={a}l}
would be the 26th of {\it Shaww\={a}l}.
On the other hand, experienced observers of the moon can deduce the number of remaining days
before a new crescent moon could be observable from the currently observed age of the moon anytime during the month.
Given that the month of {\it Shaww\={a}l} ended with the evening sunset on
AD 775 Aug 31 or Sep 1 or Sep 2, the 26th or 27th of {\it Shaww\={a}l} would then be 
AD 775 Aug 28/29 or 29/30 or 30/31 or Aug 31/Sep 1 (from sunset to sunset). And given that the sighting of the celestial object
in question here was during the morning twilight, we can date this event to a morning of AD 775 Aug 29 - Sep 1, see also Table 1. 
We cannot clarify the exact date without further information. \\
The month before {\em Shaww\={a}l} is {\em Rama\d{d}\={a}n}, the month of fasting; while one could argue that particular care
would be applied to determine the (start and) end of this month correctly (to start and end fasting neither too late
nor too early), so that the date of the beginning of {\em Shaww\={a}l} was determined more precisely than usual, it could
also well be possible that it was considered better to end {\em Rama\d{d}\={a}n} one day too late (than to stop fasting
too early), in order not to break a religious law (see also footnotes 4 and 5), 
so that {\em Shaww\={a}l} effectivelly often started one day too late. \\
We do not know how Kennedy (1990) obtained AD 775 Sep 1 as date for the end of {\it Shaww\={a}l} 158h.
Kennedy (1990) gives once {\it 26 Shaww\={a}l} and once the {\it 28 Shaww\={a}l},
but he does not comment about the date discrepancy, while Dietrich (1952) gives {\it 28 Shaww\={a}l}. \\
We would like to note that the dating uncertainty as explained above ($\pm 1$ day
due to uncertainties in crescent sighting, day counting, or moon phase determination) is often not 
taken into account when converting a Muslim calendar date to a date in the Julian or Gregorian calendar.
Some apparent inconsistencies between a given weekday and the weekday of a calculated date may
be explained by this uncertainty 
(as, e.g., for the date of the death of caliph {\it Ab\={u} Ja$^{\rm c}$far al-Man\d{s}\={u}r}, see below).
If $\pm 1$ day is the $\sim 1 \sigma$ error bar, roughly one third of such cases can even be wrong by 2 days,
e.g. Said et al. (1989) list three cases (among a total of 27 cases from AD 833 to 1513), 
where the weekday of a solar or lunar eclipse was wrong by 2 days. \footnote{They write:
{\em ... there is exact accord between the recorded and calculated dates in about two-thirds of
cases - a fairly satisfactory result} (Said et al. (1989).} With similar arguments, also Spuler \& Mayr (1961)
and Spuler (1963) argued that the conversion 
between a Muslim calendar date and the Julian or Gregorian calendar has an uncertainty of up to $\pm 2$ days,
regardless even of whether the length of the month was fixed or observed by the crescent.
An additional uncertainty of one more day could be introduced, if a month would last neither 29 nor 30 days;
while some scholars argue that this is excluded given the above Hadith (footnote 5), others could show
evidence that in a few rare cases, a Muslim month did last 31 days (see, e.g., de Blois 2000, according to
Schaefer (1992), this can happen in $0.06~\%$ of cases); however,
because this is very rare, we neglect this possibility here (it would shift the possible date by one day). \\
Both reports specify that the celestial object was visible roughly from the beginning 
until the end of twilight of that morning. For a location in or near Baghdad,
the morning twilight (for the given date range, Aug 29 - Sep 1, Julian calendar) 
is running from 4:09-4:13h (begin astronomical twilight) to 5:36-5:38h (sunrise),
for Baghdad at $33^{\circ}20^{\prime}$ north and $44^{\circ}26^{\prime}$ east,
given for (today's) time zone of Baghdad (3h east of Greenwich).
The duration of the sighting was up to some 90 min. 

\begin{figure}
{\includegraphics[angle=270,width=8.3cm]{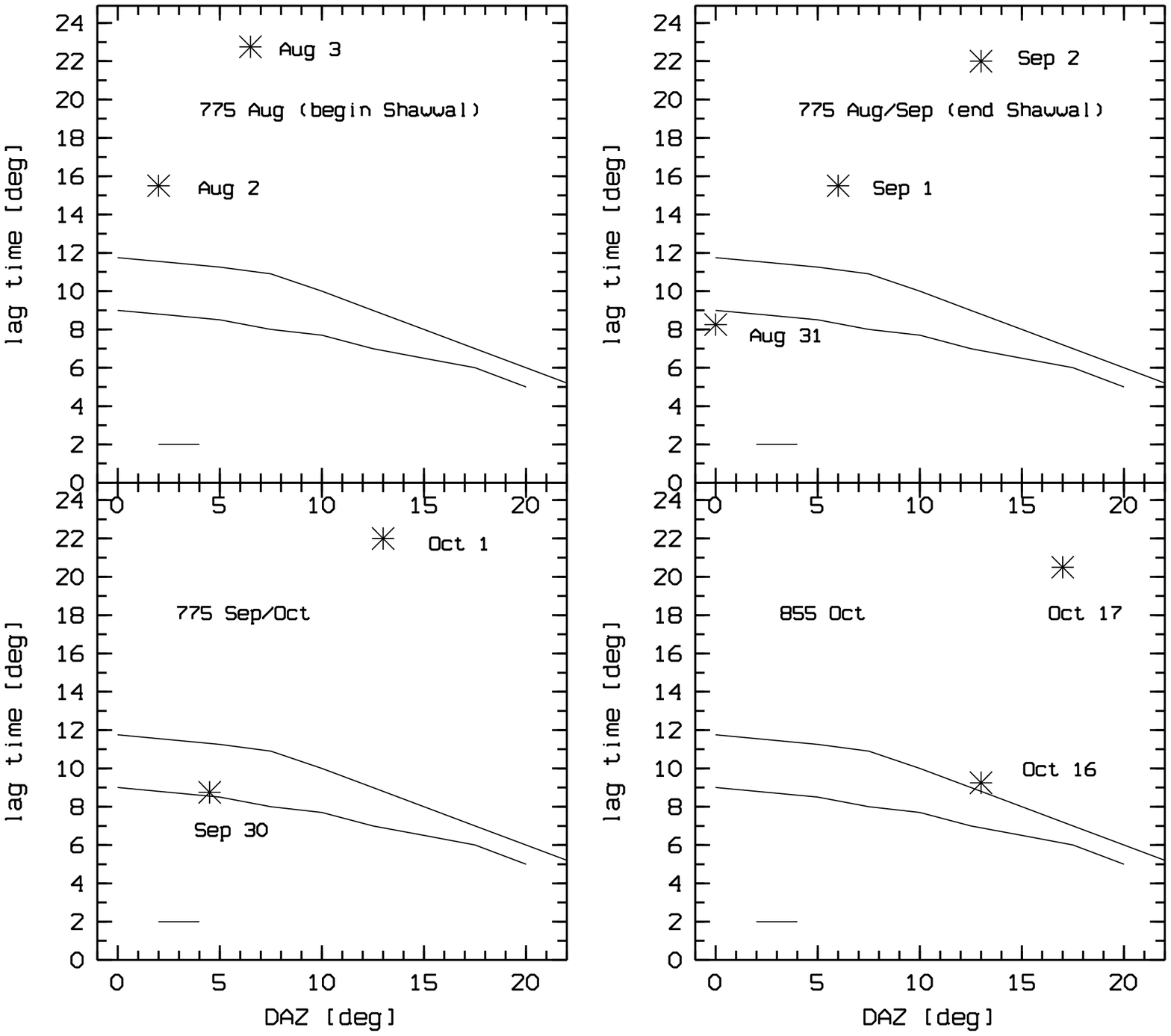}}
\caption{We plot lag-time (between sunset and moonset in degrees) versus their difference in azimut (DAZ in degrees).
We also draw two lines from Gautschy (2014) indicating the possibilty of observation:
If the moon is located clearly above the upper line, it can certainly be observed.
If the moon is located clearly below the lower line, it cannot be observed.
If the moon is located inbetween the two lines, it could possibly have been observed.
The lines from Gautschy (2014) were obtained by studying many cresent predictions and sightings
in ancient Babylon, which is not far from Baghdad, Iraq. The bar in the lower left indicates
the $\sim 2^{\circ}$ uncertainty following Gautschy (2014). We also indicate the year, month, and day of the observation,
as discussed in the text (Sect. 3, but Sect. 6 with footnote 13 for the lower right.}
\end{figure}

\begin{table}
\begin{tabular}{lll} 
\multicolumn{3}{l}{{\bf Table 1.} Date of third-to-last night in {\it Shaww\={a}l} 158h} \\ \hline
Muslim date              & \multicolumn{2}{c}{Julian calendar date (sunset to sunset)} \\ 
(sunset to sunset)       & crescent on Sep 1 & crescent on Sep 2      \\ \hline
{\it 26 Shaww\={a}l}  & Aug 28 - 29    & Aug 29 - 30 \\
\multicolumn{2}{l}{3 full nights to come}         \\ \hline
{\it 27 Shaww\={a}l}  & Aug 29 - 30    & Aug 30 - 31 \\
\multicolumn{2}{l}{with 3rd-to-last night}        \\ \hline
{\it 28 Shaww\={a}l}  & Aug 30 - 31    & Aug 31 - Sep 1 \\
\multicolumn{2}{l}{with 2nd-to-last night}        \\ \hline
{\it 29 Shaww\={a}l}  & Aug 31 - Sep 1 & Sep 1 - 2 \\
\multicolumn{2}{l}{with last night}               \\ \hline
{\it 1 Dh\={u} al-Qa$^{\rm c}$dah} & Sep 1 - 2 & Sep 2 - 3 \\ \hline
\end{tabular}
\end{table}

With one or two assumptions, we can now try to constrain the date further: \\
If the length of a Muslim month was fixed to at least 29 days and/or at most 30 days,
and if the start or end of the next or previous month would be known, we can constrain
the start or end, respectively, of the relevant month of {\it Shaww\={a}l}.
In the 2nd report about the transient event, {\it al-\d{T}abar}\textit{\={\i}} reports a little bit later,
as translated by Kennedy (1990):\\
{\em When he [caliph Ab\={u} Ja$^{\rm c}$far Al-Man\d{s}\={u}r] was in the
w\={a}d}\textit{\={\i}} {\it called Saqar, which was the last lodging on the road to Mecca, his horse stumbled and
crushed his back. He died and was buried at Bi$\,\spirituslenis{^{ }}$r Maym\={u}n ... the night when Ab\={u}
Ja$^{\rm c}$far al-Man\d{s}\={u}r died, which was Saturday, 6 Dh\={u} al-\d{H}ijjah, 158 [AD 775 Oct 7] ..}. \\
Kennedy (1990) remarks
that the weekday of the given date, AD 775 Oct 7 (6 {\it Dh\={u} al-\d{H}ijjah} 158h),
was not a Saturday as reported by {\it al-\d{T}abar}\textit{\={\i}},
but actually a Sunday - such a difference by one day could be either due to a dating error in
the report (wrong date and/or wrong weekday) or due to the one- to two-day uncertainty in the start of a new month as
set by the observation of a crescent (new) moon, see above.
According to eclipse.gsfc.nasa.gov, new moon was on AD 775 Sep 29 at 14:24h UT (Julian calendar),
so that the month of {\it Dh\={u} al-\d{H}ijjah} in the year 158h should have started in AD 775 on Oct 1 or 2:
On AD 775 Sep 30, the moon was still located too close to the Sun and the horizon, see Fig. 3;
according to Gautschy (2011), the moon was not visible before AD 755 Oct 1 at 15:11h UT at Babylon (see footnote 7),
actually first visible during the first dawn after the time given.
Hence, the month of {\it Dh\={u} al-\d{H}ijjah} probably started on Oct 1 (in case of good weather),
otherwise one day later on Oct 2 - or even on Sep 30 (in case of a false early sighting).
Hence the Muslim date of 6 {\it Dh\={u} al-\d{H}ijjah} 158h would then correspond to the day
starting at the evening sunset on AD 775 Oct 5 or 6 or 7. \\
According to Spuler \& Mayr (1961), the Muslim date of 6 {\it Dh\={u} al-\d{H}ijjah} 158h
corresponds (in the calculated Islamic calendar) to AD 775 Oct 7 (this date was also given by Kennedy 1990) - this holds
if one assumes AD 622 Jul 16 as the start of the Muslim year 1h.
Otherwise, if one would start the year 1h on AD 622 Jul 15, then the date of
6 {\it Dh\={u} al-\d{H}ijjah} 158h would correspond (in the calculated Islamic calendar) to AD 775 Oct 6.
Spuler \& Mayr (1961) also specify that AD 775 Oct 6 was a Friday and that Oct 7 was a Saturday,
while Kennedy (1990) wrote that Oct 7 would have been a Sunday. \\
In summary, we can conclude that the day 6 {\it Dh\={u} al-\d{H}ijjah} 158h (when the caliph died)
was AD 775 Oct $6 \pm 1$, with Oct 7 being a Saturday. Since {\it al-\d{T}abar}\textit{\={\i}} himself
reports that the weekday was actually a Saturday, the most likely time of the death of the caliph is the night leading to (our)
Saturday, AD 775 Oct 7 (in this case, the crescent new moon was sighted on Oct 1. \\
Since there is no evidence that this month started late, we cannot constrain the start or end of the previous month any further.

\section{\it Interpretation by the observers}

While Kennedy (1990) uses once {\em shooting star} and once {\em meteorite} for the Arabic {\em kawkab}, 
Dietrich (1952) translates it as {\em comet}. The Arabic {\em kawkab} means mainly {\em star}, but also {\em planet},
and it can also stand for any celestial object including {\em shooting star} (meteor), {\em comet},
or transient (new) star (e.g. Kunitzsch 1995).
Therefore, we translate {\em kawkab} more generally with {\em celestial object}. 
The Arabic language knows a specific word for {\em comet}, namely {\em (kawkab) dh\={u} dhanab} or 
{\em kawkab al-dhanab} meaning {\em star with a tail},
and also a specific word for {\em meteor}, namely {\em shih\={a}b} or {\em nayzak} (Kunitzsch 1995). 
The former word for meteor ({\it shih\={a}b}) is also used in the Quran -
certainly known very well to the caliph (i.e. the religious and political head 
of the Muslim community) and his group, who observed and reported the event. 
There were no specific Arabic words for bolide, nova, super-nova, kilo-nova, etc. \\
Since Dietrich (1952) translates the verb {\em inqa\d{d}\d{d}a} with {\em set}, we should consider the possibility 
that an unusual celestial object was observed while setting between the start and end of the morning twilight.
There was no particulary bright star or planet setting in the morning around the given date.\footnote{Jupiter
was setting in the south-west, but already $\sim 2.5$h after local midnight ($-2$ mag), while Moon, Venus, and Mercury 
rose shortly before the Sun.}
If the observed object would not have been unusual, it would not have been reported.
As unusual object, it could have been an unusual meteor, a comet, or some other new 
(possibly transient) object (nova-like).

If the translation of {\it inqa\d{d}\d{d}a} with {\em set} by Dietrich (1952) is correct,
the very fact that the {\em setting} was reported probably means that it was observed 
very low close to the western horizon, maybe really during its setting or close to setting.
Hence, the object location is such that it was about to set during the morning twilight
on AD 775 Aug 29 - Sep 1 (Julian calendar) as observed from the location {\em Qa\d{s}r~~$^{\rm c}$Abdaw- ayh}. 
It is well possible that it was observed and noticed just before, during, or after the 
morning dawn ({\em fajr}) prayer, which starts during the astronomical twilight and lasts until sunrise.
All prayers are to be directed towards Mecca, 
i.e. towards the south-west from any location between Baghdad and Mecca.
At this location and time, the following constellations were seen in the west:
Eri, Cet, Ari, Psc, Tri, and And - all quite distant from the
Galactic plane (where most Galactic SNe happen). Further to the west, 
some constellations crossing the Galactic plane were seen (Cas, Oph, Cyg, Lac).
However, the meaning of {\em set} for {\it inqa\d{d}\d{d}a} would be very rare and unusual.
A more correct translation is {\em fall} or maybe {\em speed down}.
In such a case, we would not be able to draw any conclusion about the direction.

In the 2nd report about the transient event, {\it al-\d{T}abar}\textit{\={\i}} reports a little bit later,
as translated by Kennedy (1990):\\
{\em According to $^{\rm c}\bar{I}$s\={a} b. Mu\d{h}ammed - 
M\={u}s\={a} b. H\={a}r\={u}n: When al-Man\d{s}\={u}r reached the last halt on the
road the Mecca, he looked inside the house he was staying in and there was written in it:
In the name of God, the Merciful, the Compassionate, Ab\={u} Ja$^{\rm c}$far, your death is drawing near, and
your years are coming to a close. There is no escape from the decree of God. Ab\={u} Ja$^{\rm c}$far, is there
a wizard or astrologer with you today who can hold back the pain of death~? ... Al-Man\d{s}\={u}r then
dictated the two verses ... He [Ab\={u} Ja$^{\rm c}$far al-Man\d{s}\={u}r] gave orders to move from the lodging,
seeing what had occurred as an evil omen, and he rode off on horseback. When he was in the
w\={a}d}\textit{\={\i}} {\it called Saqar, which was the last lodging on the road to Mecca, his horse stumbled and
crushed his back. He died and was buried at Bi$\,\spirituslenis{^{ }}$r Maym\={u}n ... the night when Ab\={u}
Ja$^{\rm c}$far al-Man\d{s}\={u}r
died, which was Saturday, 6 Dh\={u} al-\d{H}ijjah, 158 [AD 775 Oct 7] ..}.\footnote{See Sect. 3,
where we found that the death indeed most certainly took place on Saturday, AD 775 Oct 7.}

The above report can be interpreated as follows:
The caliph thought to have seen the reported text written inside the house;
in reality he reports about his own thoughts and considerations, 
namely that he may be close to his death (due to his illness) 
and that he is sorry not to have an astrologer with him,
who could have tried to interprete what has happened and what is seen as evil omen
by the caliph ({\em seeing what had occurred as an evil omen}).
It is possible that the caliph was worried about the sigthing of the
transient celestial object and its possible (negative) astrological meaning. 

\section{\it A bolide~?}

Given the use of the word {\em kawkab} as star or meteor and the translation by
Kennedy (1990) as {\em a shooting star [kawkab] fell [{\it inqa$\d{d}\d{d}$a}]}, 
it is certainly possible to interprete this observation as meteor or bolide,
whose trail or trace ({\em athar}) was visible for tens of minutes, 
roughly from the beginning to the end of twilight. Such a scenario would
be possible, but still unusual (and not understood), so that it was reported.
Persistent train emission of a bolide or fireball remain in the path and can last for up to some $\sim 30$ 
minutes - due to recombination of oxygen catalysed by Na and FeO - later distorted by 
atmospheric winds.\footnote{www.leonid.arc.nasa.gov/meteor.html}

However, if the sighting of a bolide trail would be reported as or connected with a {\em meteor},
do we need to expect the meteor to have been observed as well~?
If yes, then we would be surprised that the Arabic word {\it shih\={a}b} is not used here
(as it is in the Quran for meteors, of course well-known to the caliph and his group).
Here, we also have to consider the so-called {\em Sternschnuppenmythos}, 
a myth in the Quran about meteors:
Satan(s), who try to spy on angels, can be driven away by throwing
a {\it shih\={a}b} (meteor) at them (Kunitzsch 1989, 1995), 
and a ritual action during the pilgrimage 
in Mecca includes throwing stones at a (virtual) satan (to signify defiance of satan):
Since the caliph and his group were on their way to Mecca for pilgrimage ({\it al-\d{T}abar}\textit{\={\i}}),
the Sternschnuppenmythos is relevant here.
In the Quran (translation by Asad (1980), our additions in round brackets (), his additions in square brackets []),
the following text is given: \\
Sura 15, 16-18: {\em 16 And, indeed, We have set up in the heavens great constellations, and endowed them with beauty for all to behold; 
17 and We have made them secure against every satanic (shay\d{t}\={a}n) force accursed, 18 so that anyone who seeks to learn [the unknowable] by stealth 
is pursued by a flame (meteor, {\it shih\={a}b}) clear to see.} \\
Sura 37, 6-10: {\em Behold, We have adorned the skies nearest to the earth with the beauty of stars, 7 and have made them secure 
against every rebellious, satanic force, 8 [so that] they [who seek to learn the unknowable] should not be able to overhear 
the host on high [the angelic forces, whose "speech" is a metonym for God's decrees], 
but shall be repelled from all sides, 9 cast out [from all grace], with lasting suffering in store for 
them [in the life to come]; 10 but if anyone does succeed in snatching a glimpse [of such knowledge], he is [henceforth] pursued by a piercing flame
(meteor, {\it shih\={a}b}).} \\
Sura 67, 5+6: {\em And, indeed, We have adorned the skies nearest to the earth with lights [stars], 
and have made them the object of futile guesses for the evil ones [from among men]: 
and for them have We readied suffering through a blazing flame
6 for, suffering in hell awaits all who are [thus] bent on blaspheming against their Sustainer: and how vile a journey's end!} \\
Sura 72, 8+9: {\em And [so it happened] that we reached out towards heaven: 
but we found it filled with mighty guards and flames (meteors, {\it shuhub},
the plural form of {\it shih\={a}b}),
9 notwithstanding that we were established in positions 
[whi- ch we had thought well-suited] to listening to [whatever secrets might be in] it
and anyone who now [or ever] tries to listen will [likewise] find a flame (meteor, {\it shih\={a}b}) lying in wait for him!} \\
These texts are interpreted as indication that one must not use non-ethical (i.e. satanic) means (including astrology) 
to obtain knowledge about the future (spying on angels), e.g. Asad (1980) writes: {\em The statement that God 
has made the heavens secure against such satanic forces obviously implies that He has made it impossible for the latter to obtain, 
through astrology or what is popularly described as occult sciences, any real knowledge of that which is beyond the reach of human 
perception.}

A possible (though speculative) conclusion here is that the caliph and his group would probably have used
the word {\it shih\={a}b}, if they would have seen a meteor connected with the transient event.
Partly because they did not use that word we have to consider an alternative interpretation (Sect. 6).

Alternatively, the word could have changed during the (at least partly oral) transmission
from the observers to {\it al-\d{T}abar}\textit{\={\i}}. Also, it might be possible that the observers did not call the event/object
{\em meteor} (or {\it shih\={a}b}), because what they observed was not a normal meteor, which should have been
visible for only $\sim 1$ second (normal meteor), but was visible for much longer (bolide as meteor with trail).

\section{\it A nova-like event~?}

It is important to compare {\it al-\d{T}abar}\textit{\={\i}}'s report and wording to Arabic reports about historic SNe.

There are a few Arabic reports about SN 1006 (Goldstein 1965):
{\it $^{\rm c}$Al}\textit{\={\i}} {\it ibn Ri\d{d}w\={a}n} uses the words {\em athar} and {\em nay- zak},
which mean {\em trace} (or maybe {\em rays}) and {\em spectacle}, respectively 
(Goldstein 1965).\footnote{Based on the ecliptic longitudes of planets, moon, sun, and lunar node also given in the 
report of {\it $^{\rm c}$Al}\textit{\={\i}} {\it ibn Ri\d{d}w\={a}n}, it was possible to calculate the observing time (AD 1006 Apr 30) and the
ecliptic longitude of the SN itself (with error bar from {\it $^{\rm c}$Al}\textit{\={\i}} {\it ibn Ri\d{d}w\={a}n}'s apparent measurement precision);
together with the declination limit from the St. Gallen observation and the Chinese right ascension range
(from Chinese {\em lunar lodge}), it was then possible to identify the SN remnant (e.g. Stephenson \& Green 2002).}
{\it Ibn al-Ath}\textit{\={\i}}{\it r} and {\it Ibn al-Dja- wz}\textit{\={\i}} both wrote about a large {\em kawkab} (large star) and its rays 
(Goldstein 1965). An anonymous report from Mauretania mentions a {\em great star} [{\it nadjm}] {\em among the comets}
and a {\em nay- zak} (Goldstein 1965). The word {\em nayzak} is often
translated as {\em comet}, but it can also mean just something like
{\em bright transient celestial object}, because at the time of SN 1006 (and SN 1054),
Arabic scholars did not know about transient celestial phenomena like novae, super-novae,
or even optical transients due to GRBs or kilo-novae.
Hence, they used the words they had for other, known transient celestial (or presumably
atmospheric) objects like comets and meteors (while real stars were considered constant
in brightness following Aristotle).

There is also one report about SN 1054 (Brecher et al. 1978):
{\it Ibn Ab\={\i} U\d{s}aybi$^{\rm c}$a} quoting {\it Ibn Bu\d{t}l\={a}n} writes about the {\em athar}\textit{\={\i}} {\it kawkab} as
{\em the star leaving traces} or {\em spectacle star} (Brecher et al. 1978).
Regarding the translation of {\em athar}, Brecher et al. (1978) write:
{\em ... a novel astronomical or meteorological phenomenon ... characterised by its
transient, explosive or spectacular appearance. Apparent star, phenomenon or
spectacle might be equally viable translations}.

Given that Arabic reports about SNe 1006 and 1054 have used the words
{\em kawkab} and {\em athar}, just as in the {\it al-\d{T}abar}\textit{\={\i}} report for AD 775,
we should continue to consider a nova-like interpretation.
The word {\em athar} (for {\em trace}) in our text is in the form of {\em atharuhu} meaning {\em its trace}.

According to {\it al-\d{T}abar}\textit{\={\i}}, the object was observed only during the morning twilight
of AD 775 Aug 29 - Sep 1 (Julian calendar), there are no positive nor negative reports about the previous or following days or nights.
There are also no reports about celestial objects at that time by other sources like
European or eastern Asian observers. If it were a more or less normal nova (or even a SN),
it should have been visible for several weeks (or several months, as SN), unless it would have been relatively
faint such as several mag (so that it was not noticed in other nights).
However, the fact that this object was detected during twilight, shows that it was brighter than
about $-0.3$ mag, the naked-eye limit around twilight (Schaefer 1993).
On the other hand, {\em it remained visible until sunrise}, so that it was not visible
during the day-time; hence, it was fainter than about $-3$ mag, the day-time naked-eye limit (Schaefer 1993).
Hence, the brightness was inbetween $-3$ and $-0.3$ mag.
If it was really setting during or soon after the morning twilight, it may have been above the horizon
for the whole night or large parts of it.
Therefore, it may have been visible for up to one night, maybe less than a few hours towards 
the constellations Eri, Cet, Ari, Psc, Tri, and And, which were visible in the west. 
(We cannot assume that the (non-professional) observers did observe the whole night, so that we cannot
exclude that the object was visible the whole night. However, if it would have been very
bright already in the first half of the night in Arabia, it should have been observed by the Chinese,
but there are no such reports.)
Kilo-novae due to short GRBs are expected to be visible for only a few hours (Metzger \& Berger 2012; Piran, Nakar, Rosswog 2013).
Both the brightness and the short time-scale of the fading (up to one day) as reported by {\it al-\d{T}abar}\textit{\={\i}} are
not inconsistent with a kilo-nova expected after a short GRB (HN13).
As kilonova with peak brightness of about $-3$ to $0$ mag, it would be at $\sim 3$ to 9 kpc.

Finally, we should consider which Arabic words were used otherwise by {\it al-\d{T}abar}\textit{\={\i}} in his chronicle 
for meteors and the {\em falling-down} of meteors, in particular close in time to the AD 775 report. 
For the year 241h (AD 855/856), we found in {\em The History of al-\d{T}abar}\textit{\={\i}}, as translated by us: \\
{\em In this year there sped down (inqa\d{d}\d{d}at) the stars (al-kaw\={a}k- ib, plural of {\em kawkab}) in Baghdad 
and became scattered, in the night of Thursday 1 Jum\={a}d\={a} II.} \\
According to eclipse.gsfc.nasa.gov, new moon was on AD 855 Oct 14 at 21:50h UT in the Julian calendar ($\pm 2$ min according to Morrison \& Stephenson 2004).
so that the 1st day of the new month ({\em Jum\={a}d\={a} II}) probably started on the evening of AD 855 Oct 16 or 17 (maybe already Oct 15 
in case of an extremely early sighting), see Fig. 3.
According to Spuler \& Mayr (1961), the 1st of {\em Jum\={a}d\={a} II} in 241h was AD 885 Oct 17 
and indeed a Thursday (since midnight).\footnote{Kennedy (1990) gives AD 855 Oct 17,
but an incorrect weekday (Tuesday), while the Arabic text clearly says {\em al-Kham}\textit{\={\i}}{\it s}, i.e. Thursday; 
this mistake by Kennedy (1990) could be due to an eye-slip, namely by reading off the weekday of 1 {\em Jum\={a}d\={a} I} 
of 241h (a Tuesday) instead of 1 {\em Jum\={a}d\={a} II}.} \\
The verb used in the original Arabic here was {\em inqa\d{d}\d{d}at} and the celestial objects
were called {\em al-kaw\={a}kib} (plural of {\em kawkab}), meaning {\em stars falling down}. \\
As noticed before (Newton 1972, Dall'Olmo 1978), this meteor storm was also observed in Europe. 
Annales Fulden- ses, a chronicle of the Christian monastery in Fulda, Germany, reports: 
{\em On the 17th of October, densely packed small fires (igniculi) like lances (spiculorum) were zipping 
towar- ds west through the air during the whole night}; this is our translation following the text in Rau (1960), 
who gives the original Latin\footnote{Mense vero Octobris XVI Kal. Novembr, per totam noctem igniculi instar spiculorum 
occidentem versus per aerem densissime forebantur.} together with a German translation.\footnote{Dall'Olmo (1978)
remarks that Emperor Lothar would have {\em prayed neglecting all his affairs} because of this sighting according to
the Fulda chronicle; however, the Annales Fuldenses report after the meteor sighting that Emperor Lothar would have entered
a monastery and died there, not combining this with the meteor sighting at all; in addition, Emperor Lothar died
already on AD 855 Sep 29 (according to, e.g., Annales Bertinini), hence before the meteor sighting.} 
The Fulda chronicle reports about the time AD 714 to 902 and was probably written by three authors: Einhard from AD 714 to 838 
(partly based on other sources), but including eyewitness reports for the last few decades, then from AD 838 to 863 by the monk 
Rudolf of Fulda (died 865), who is then possibly the eyewitness for the meteor sighting, and the final part by Meginhard, 
another monk from Fulda, the latter both eyewitnesses; alternatively, one author may have compiled the work since AD 882; 
six copies are available (Rau 1960). 
The dating of astronomical events is quite precise in Annales Fuldenses:
Original reports about solar ecli- pses are given for AD 840 May 5 and 878 Oct 29 (Newton 1972),
which were actually exactly on those days and visible in western Europe (eclipse.gsfc.nasa.gov/solar.html);
an eclipse reported for AD 817 Feb 5 as {\em solar eclipse} 
was a lunar eclipse on AD 817 Feb 5 visible in western Europe (eclipse.gsfc.nasa.gov/lunar.html);
and then original reports about lunar eclipses are given for AD 842 Mar 30 and 878 Oct 15 (Newton 1972),
which were actually exactly on those days and visible in western Europe (eclipse.gsfc.nasa.gov/ lunar.html).  \\
This additional, completely independent report can be seen not only as confirmation for the meteor sighting itself,
but also for the date: Even though the Fulda report gives Oct 17 as date for the sighting ({\em during the whole night}), 
it would remain uncertain a priori whether it means the night Oct 16/17 or 17/18. However, since the night of Oct 17/18
would be a date corresponding to Oct 18 in the Islamic calendar, which is excluded from the Arabic report (AD 855 Oct $16 \pm 1$), 
we can conclude that the observations were made in the night Oct 16/17. This can then constrain the first sighting of the 
new lunar crescent to the evening of AD 855 Oct 16 (1 Jum\={a}d\={a} II); and it can confirm the week-day given by {\it al-\d{T}abar}\textit{\={\i}}.  \\
For a meteor storm on AD 855 Oct $16 \pm 1$ in the Julian calendar (which would correspond to AD 855 Oct $20 \pm 1$ on the Gregorian calendar),
it could have been the Orionids (peaking Oct 21 in the Gregorian calendar). \\
From these texts, we can see that {\it al-\d{T}abar}\textit{\={\i}} did use {\em al-kaw- \={a}kib} 
together with the verb {\em inqa$\d{d}\d{d}$a} for {\em stars falling down} or meteors (instead of {\em shih$\bar{a}$b} for meteor as in the Quran). 
Hence, the interpretation of the celestial event in AD 775 as bolide is not problematic in this regard. 

Without any further information, we cannot conclude that this observation relates to a nova-like event.
The interpretation as bolide, though, does not pose similar problems.

\section{\it Summary}

In his two similar reports about an unusual transient celestial object in AD 775, the Arabic historian {\it al-\d{T}abar}\textit{\={\i}}
uses the words {\em kawkab} and {\em atharuhu}, i.e. two of the three words used by Arabic authors in their reports about SNe
1006 and 1054. The event can be dated to AD 775 Aug 29 - Sep 1 (Julian calendar), observed by caliph {\it al-Man\d{s}\={u}r}
and/or member(s) of his group - observed near {\it Qa\d{s}r~~$^{\rm c}$Abdawayh} in the Baghdad area (on the way from Baghdad to Mecca).
It seems that the object was observed only for up to some 90 min during the morning twilight,
hence it was quite bright (between $-3$ and $-0.3$ mag) and possibly fading fast. 
We cannot exclude that this report relates to an observation
of an optical transient (kilo-nova) expected in connection to a short GRB.
Such an event was suggested as cause for the $^{14}$C variation around AD 774/5 by HN13. 
However, since the word {\em kawkab} can also be used for meteor, it is also possible to interprete
the observed object as the trace ({\em athar}) of a bolide seen for some tens of minutes.
Since the verb {\em inqa$\d{d}\d{d}$a} normally means {\em fall}
or in this context also {\em fade}
(instead of {\em set} as used by Dietrich 1952), we prefer the interpretation as bolide.
A bolide observed in AD 775 between Aug 29 and Sep 1 in the Julian calendar (which corresponds to AD 775 Aug 31 - Sep 3 in the Gregorian calandar),
could have belonged to either the Gamma Doradids (peaking 28 Aug) or the Alpha Aurigids (peaking 31 Aug) meteor showers. 
The previous translation by Dietrich (1952), namely {\em a comet [kawkab] had set}, is not correct. \\
We have discussed in detail as to how to convert a Muslim calendar date to the Julian (or Gregorian) calendar by first using the
calculated Muslim calendar ($\pm$ at least 1 or 2 days) and then by investigating, when exactly the new crescent moon was visible at the relevant location.
The sighting of the crescent moon can be complicated or postponed by bad weather, high extinction at low latitude, or difficult landscape,
but also false early sightings are possible. For the meteor storm dated AD 855 Oct 17 by Annales Fuldenses, we could show that
it corresponds exactly to the date and week-day given by {\it al-\d{T}abar}\textit{\={\i}} as Thursday {\it 1 Jum\={a}d\={a} II} in the year 241h.
We could also confirm that the week-day given by {\it al-\d{T}abar}\textit{\={\i}} for the death of caliph {\it Ab\={u} Ja$^{\rm c}$far al-Man\d{s}\={u}r},
namely Saturday, {\it 6 Dh\={u} al-\d{H}ijjah} 158h, can be correct, given the lunar date line for that lunar month
(while Kennedy 1990 considered the week-day given by {\it al-\d{T}abar}\textit{\={\i}} to be wrong).

\acknowledgements
We would like to thank H. \"Ozbal for discussion about tree rings.
We are also grateful to W. Rada and T. Seidensticker for advise regarding Muslim calendar rules.
We would also like to thank S. Leder and T. Seidensticker for information about the work of {\it al-Haytham ibn $^{\rm c}$Ad}\textit{\={\i}}.
T. Seidensticker also commented on an earlier version of this manuscript.
We also acknowledge the moon phase as well as solar and lunar eclipse predictions by Fred Espenak, NASA/GSFC, on eclipse.gsfc.nasa.gov,
as well as moon phase and visibility predictions from R. Gautschy on www.gautschy.ch/~rita/archast/mond/Babylonerste.txt.
RN is also grateful to Dagmar L. Neuh\"auser for intensive discussion about both the AD 774/5 $^{14}$C event and the celestial event reported here.
RN also acknowledges discussion with M. Mugrauer about bolid- es.
RN would like to thank F. Sezgin, G. Yildiz, M. Amawi, and C. Ehrig-Eggert (Institut f\"ur Geschichte der Arabisch-Islamischen Wissenschaften, Frankfurt) 
as well as B. Dincel for their help with searching Arabic chronicles.
Finally, we would also like to thank our referee, Prof. H. Basurah, for several good suggestions.

{}

\end{document}